\documentclass[authoryear,final,3p,times,twocolumn]{elsarticle}

\usepackage{epsfig}

\usepackage{natbib}

\usepackage{amssymb}

\usepackage[bookmarks = true, bookmarksnumbered = true, pdfpagemode =None, pdfstartview = FitH, pdfpagelayout = SinglePage, colorlinks = true, urlcolor = blue, citecolor = blue]{hyperref}

\begin{document}
\begin{frontmatter}
\title{Exploring the Diversity of Jupiter-Class Planets}

\author[ox]{Leigh N. Fletcher}
\ead{fletcher@atm.ox.ac.uk}
\author[ox]{Patrick G.J. Irwin}
\author[ast]{Joanna J. Barstow}
\author[sron]{Remco J. de Kok}
\author[zur]{Jae-Min Lee}
\author[ast]{Suzanne Aigraine}

\address[ox]{Atmospheric, Oceanic \& Planetary Physics, Department of Physics, University of Oxford, Clarendon Laboratory, Parks Road, Oxford, OX1 3PU, UK}
\address[ast]{Department of Physics, University of Oxford, Denys Wilkinson Building, Keble Road, Oxford OX1 3RH, UK}
\address[sron]{SRON Netherlands Institute for Space Research, Sorbonnelaan 2, 3584 CA Utrecht, The Netherlands}
\address[zur]{Institute for Theoretical Physics, University of Zurich, CH-8057, Zurich, Switzerland}

\begin{abstract}
\textbf{Contribution to a Royal Society Theo Murphy Discussion Meeting (2013) `Characterizing exoplanets: detection, formation, interiors, atmospheres and habitability'.}

Of the 900+ confirmed exoplanets discovered since 1995 for which we have constraints on their mass (i.e., not including Kepler candidates), 75\% have masses larger than Saturn (0.3$M_J$), 53\% are more massive than Jupiter, and 67\% are within 1 AU of their host stars.  When Kepler candidates are included, Neptune-sized giant planets could form the majority of the planetary population.  And yet the term `hot Jupiter' fails to account for the incredible diversity of this class of astrophysical object, which exists on a continuum of giant planets from the cool jovians of our own solar system to the highly-irradiated, tidally-locked hot roasters.  We review theoretical expectations for the temperatures, molecular composition and cloud properties of hydrogen-dominated Jupiter-class objects under a variety of different conditions.  We discuss the classification schemes for these Jupiter-class planets proposed to date, including the implications for our own Solar System giant planets and the pitfalls associated with compositional classification at this early stage of exoplanetary spectroscopy.  We discuss the range of planetary types described by previous authors, accounting for: (i) thermochemical equilibrium expectations for cloud condensation and favoured chemical stability fields; (ii) the metallicity and formation mechanism for these giant planets; (iii) the importance of optical absorbers for energy partitioning and the generation of a temperature inversion; (iv) the favoured photochemical pathways and expectations for minor species (e.g., saturated hydrocarbons and nitriles); (v) the unexpected presence of molecules due to vertical mixing of species above their quench levels; and (vi) methods for energy and material redistribution throughout the atmosphere (e.g., away from the highly irradiated daysides of close-in giants).   Finally, we will discuss the benefits and potential flaws of retrieval techniques for establishing a family of atmospheric solutions that reproduce the available data, and the requirements for future spectroscopic characterisation of a set of Jupiter-class objects to test our physical and chemical understanding of these planets.
\end{abstract}
\end{frontmatter}

%\linenumbers

\section{The Jupiter Classification}

The high-temperature hydrogen-rich atmospheres of extrasolar giant planets (EGPs) in close orbits around their parent stars have made them ideal candidates for preliminary spectroscopic characterisation.  Giant exoplanets (i.e., Neptune sized and larger) appear to be commonplace:  from a catalog of confirmed exoplanets with mass determinations\footnote{http://exoplanet.eu}, 75\% of all planets discovered to date have masses larger than Saturn (0.3$M_J$), 53\% are more massive than Jupiter, and 67\% are within 1 AU of their parent star.  However, this could simply be the result of observational bias for the larger giants - when the list of Kepler candidate objects is included \citep{13fressin, 13batalha}, Neptune-sized giants could comprise a significant percentage \citep[30\% or larger,][]{13fressin} of planetary objects beyond our solar system, restricting Jupiter-sized objects to 5\% or less.   This article will focus on the taxonomy of hydrogen-rich gaseous exoplanets, in particular the Jupiter-class with radii $>6R_E$ \citep[following the prescription of the Kepler team,][]{13fressin}, to which our own Jupiter (11 $R_E$) and Saturn (9.1 $R_E$) belong.  However, many of the conclusions may be equally valid for Neptune-class objects ($2-6R_E$) or hydrogen-rich super earths (1.25-2 $R_E$) that may dominate the planetary populations beyond our solar system.  The conditions revealed on these worlds are significantly different from the gas giants of our own solar system, and yet these hot ($\approx2500$ K) and cold ($\approx100$ K) jovians must exist on a continuum of planetary types that can be categorised in terms of a range of atmospheric phenomena.  

Transit spectroscopy of a handful of EGPs tentatively revealed the presence of simple molecules of hydrogen, carbon and oxygen \citep[water,methane, CO and CO$_2$,][]{07tinetti,08swain,09swain, 11gibson}, along with sodium \citep{02charbonneau}, atomic H and other neutral species in their upper atmospheres.  Some of these conclusions have been refined and questioned in subsequent years, and observers have gone to great lengths to confirm or refute these molecular detections to move the field forward.  Atmospheric models have revealed the importance of strong stellar insolation, the possible presence of atmospheric hazes \citep[e.g.,][]{08pont}, stratospheric thermal inversions \citep{09swain}, atmospheric winds \citep{10snellen} and longitudinal temperature contrasts \citep{07knutson}.  Approximately 25-30\% of the confirmed planets are known to transit their host stars, permitting spectroscopic characterisation via the transit method biased towards highly-irradiated, close-in giant planets on short period orbits.  A considerably smaller number of directly-imaged planets well separated from their host stars on longer orbits (young and hot worlds) are presently available for spectroscopic characterisation, but it is hoped that this number could increase in the near future.   So, given the sparse information we have on the composition, dynamics and chemistry of these EGPs, development of classification systems may seem premature.  But they serve a useful purpose, allowing us to bring a degree of order to the patterns in the emerging pantheon of planetary types being discovered today, providing a vernacular for their discussion \citep[by analogy to the categorisation of brown dwarfs,][]{05kirkpatrick} and testable hypotheses to be addressed by future spectroscopic missions. 

A series of one- and two-dimensional classification systems have been proposed in the decade since the work of \citet{03sudarsky}, who used a solar composition and thermochemical equilibrium to study the influence of stellar irradiation and its implications for atmospheric composition and cloud formation.  Put simply, the higher the irradiation, the hotter the atmosphere and the more refractory species are released from the condensed phase to interact with the emission from the planetary photosphere (broadly speaking, the middle atmosphere of the planet from the stratosphere to the upper troposphere).  At the highest temperatures, oxides of titanium and vanadium would be available \citep[if the planet is oxygen-rich,][]{12madhu} to serve as strong UV/visible absorbers, generating stratospheric temperature inversions \citep[e.g.,][]{03hubeny} and leading \citet{08fortney} to develop a one-dimensional classification system (pM and pL classes, by analogy to M and L brown dwarfs) based on the presence or absence of a thermal inversion.  As we shall see in Section \ref{spectra}, existing observations are generally not robust enough to confirm the stratospheric temperature gradients, but several studies have identified highly-irradiated EGPs with no thermal inversions at all\citep{11madhu}, betraying the simplicity of any irradiation-driven categorisation scheme.

% Blecic (2011) and Anderson (2011) also reveal absent stratospheres, but these are unpublished.

Indeed, Section \ref{processes} reveals several other factors influencing the atmospheric temperature and composition, and good reviews of these processes can be found in \citet{10seager} and \citet{10burrows}.  Gravitational settling and cold-trapping (the same process that keeps the Earth's stratosphere free of water vapour) may deplete TiO and VO from EGP atmospheres and limit their role in forming stratospheric inversions \citep{09spiegel}.  Furthermore, an oxygen-poor, carbon-rich atmosphere would limit the production of these oxides, as well as having dramatic effects on other carbon species \citep{09helling, 11madhu, 12madhu}.  Chromospheric activity of the parent stars (e.g., flares and cosmic rays) can either inhibit photochemistry and destroy potential UV/visible absorbers, or excite additional ion chemistry due to excess charged particle bombardment, which can generate upper atmospheric hazed that contribute to the radiative budget \citep{09zahnle_soot, 10knutson, 13moses}, although the residency times and importance of these species for generating thermal inversions is uncertain.  Other potential stratospheric absorbers may be tied up in condensed phases not yet considered in models.  Different thermal structures will affect an atmosphere's susceptibility to photochemistry and vertical mixing \citep{10line, 11moses, 13moses}. Vertical and horizontal mixing by eddy diffusion, convection and wave propagation would also be responsible for moving energy and material from place to place, causing the composition to deviate from the expectations of equilibrium \citep[e.g.,][]{10line,11moses,10showman}.  Each of these processes could provide additional dimensions to a classification scheme for EGPs.  Of particular note is the recent two-dimensional scheme devised by \citet{12madhu} and \citet{13moses}, which uses both the stellar irradiance and the chemical dependence on the C/O ratio, described below.

\begin{figure*}[tbp]
\centering
\includegraphics[width=16cm]{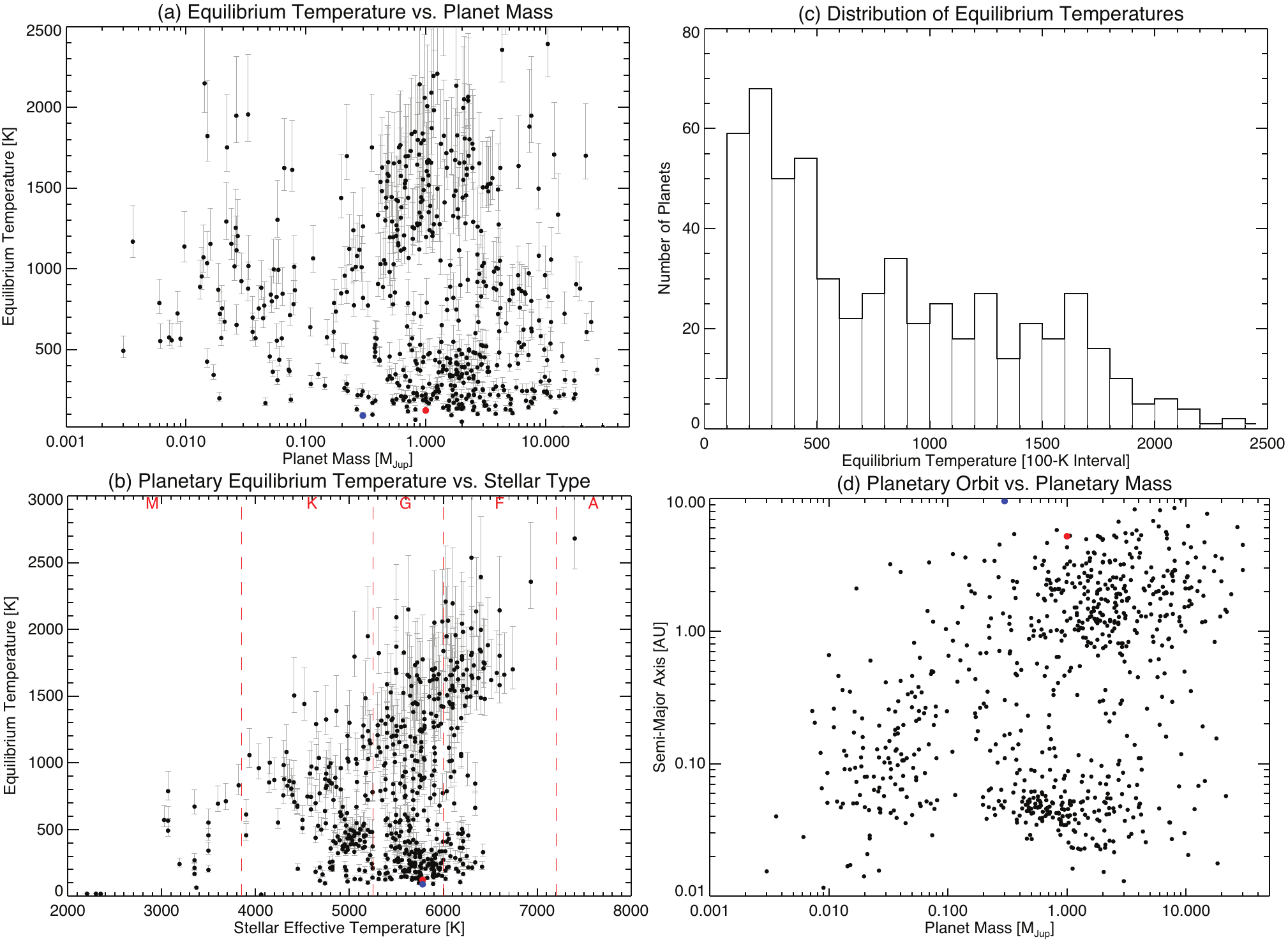}
\caption{The diversity of Jupiter-class exoplanets discovered to date with constrained masses (all data from the catalogue hosted at exoplanet.eu).  Panel (a) plots the equilibrium temperature against planetary mass (with Jupiter as a red circle and Saturn as a blue circle for comparison); (b) shows the range of equilibrium temperatures discovered for each stellar type from F to M; and (c) shows the number of planets discovered to be occupying a particular equilibrium temperature range.  Although $T_{eq}$ is a poor proxy for the true atmospheric temperatures, it does permit a first-order attempt at classifying these EGPs.  Error bars depict the range of $T_{eq}$ for bond albedoes between 0.0 and 0.3, and the difference between purely dayside re-emission for full energy redistribution).  Panel (d) shows the planetary mass and orbit of all exoplanets discovered to date.   }
\label{Teqm}
\end{figure*}

Although the true atmospheric temperature-pressure ($T(p)$) profile is desirable for a complete discussion of chemistry and dynamics, in reality we must use a proxy, the equilibrium temperature $T_{eq}$ based on the stellar luminosity, $L$; the bond albedo $a$; the degree of horizontal temperature redistribution $f$; the orbital distance $r$ and Stefan-Boltzmann constant, $\sigma$ \citep{03sudarsky}:
\begin{equation}
T_{eq}=\left[ \frac{L(1-a)}{16\pi\sigma r^2f}\right]^{1/4}
\end{equation}
In Fig. \ref{Teqm}, we estimate $T_{eq}$ for all planets discovered to date, accounting for the uncertainty in the bond albedo (from zero to a Solar-System-like value of 30\%) and the efficiency of redistribution ($f=1$ for a full redistribution of incoming flux; $f=0.5$ if radiation is only emitted from the dayside atmosphere).  This differs from the effective temperature ($T_{eff}$) which also takes into account an internal boundary flux that varies according to the age and thermal history of the planet, such as the excess luminosity of the giant planets in our own solar system \citep{91pearl}.  The number of planets discovered in a particular $T_{eq}$ category is also presented in Fig. \ref{Teqm}.  Although a poor proxy for the true $T(p)$, this quantity does at least permit preliminary categorisations \citep[e.g.,][]{03sudarsky,08fortney,12madhu}, and will be used as a guide in the text that follows.

\section{Atmospheric Processes}
\label{processes}

\begin{figure*}[tbp]
\centering
\includegraphics[width=13cm]{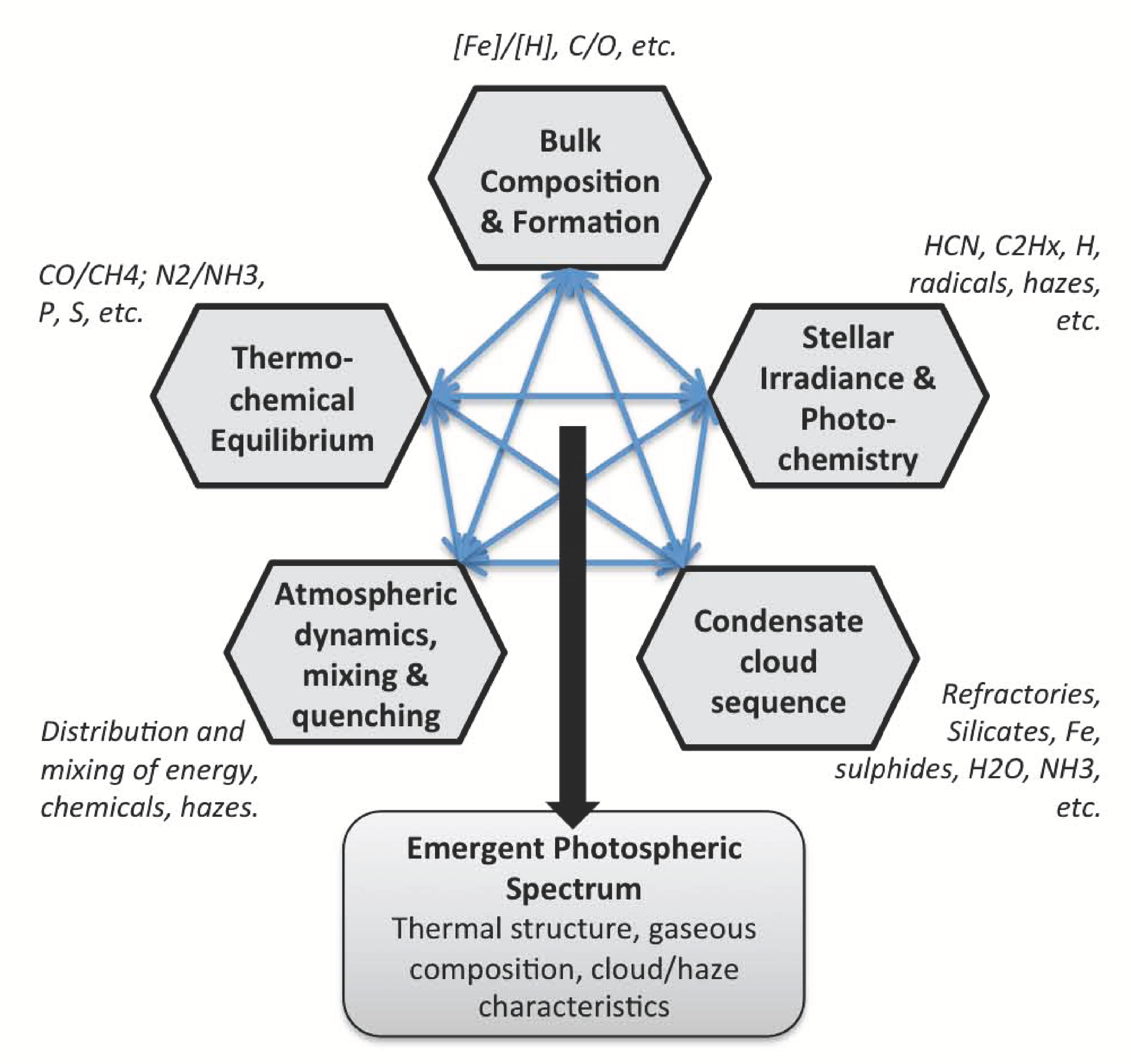}
\caption{Schematic showing the principle atmospheric processes governing the photospheric compositions of the family of Jupiter-class exoplanets, and hence shaping the emergent spectrum measured in transits and eclipses. The arrows signify that these processes are all closely coupled - the details of the condensate sequence, for example, depends strongly on the availability of source material to form clouds, and hence on the bulk composition imprinted at the time of planet formation.}
\label{cartoon}
\end{figure*}

In the following sections we review some of the processes responsible for shaping the composition, and hence the emergent spectra and potential multi-dimensional classification schemes, of the Jupiter-class exoplanets (Fig. \ref{cartoon}).  

\subsection{Planetary Origins}

The zoo of species available for atmospheric chemistry and cloud formation are initially determined by the bulk composition of a planet, imprinted during the earliest accretion stages \citep[see the recent review by][]{10lodders}.  Different formation mechanisms, whether from direct collapse of gases from the protoplanetary disc \citep[the disc-instability model,][]{97boss, 01boss} or from the formation of rock-ice cores prior to runaway gas accretion \citep[the core accretion model,][]{80mizuno, 96pollack} will imprint a different balance of chemical elements and isotopic ratios on this bulk composition. Further complicating matters, the availability of materials to use as planetary building blocks \citep[metals, oxides, silicates, sulphides and ices,][]{10lodders} varies with position in the disc, from refractories that persist at the high temperatures of the inner nebula, to the ices of water, methane, ammonia and other simple molecules which are common in the cold outer nebula.  Modelling these radial distributions of refractories and volatiles requires knowledge of the parent star composition (e.g., its metallicity, [Fe]/[H], imprinted by its own protostellar nebula), the radial temperature and pressure structure in the disc, the degree of turbulent mixing within the disc, and the stability and importance of a sequence of `snow lines' where key volatiles condense, of which water ice is the most important but other species can play a role \citep[e.g., CH$_4$/CO, N$_2$/NH$_3$, etc.,][]{10lodders, 11kavelaars, 11oberg}.  Finally, a planet's bulk composition will depend on the timing and location of planet formation within a disc that cools with age; subsequent migration of the planet through the disc; and the method of delivery of the heavier elements (direct collapse of the gas or as in-falling planetesimals enriched in certain species).

Even within our own solar system, this imprint of the chemical conditions of planetary formation is hard to test unambiguously from remote sensing, and open to many different interpretations.  Although core-accretion and subsequent migration is generally accepted as the explanation for the current distribution of planets in our solar system \citep[e.g.,][]{99owen, 05gautier, 05tsiganis}, the use of atmospheric remote sensing to support this theory relies on interior models to relate the composition of the envelope to the planetary interior and core.  If the mixing of the heavy elements that formed the original planetary core into the upper atmospheric envelope is inefficient, then the observed atmospheric heavy-element enrichment will not necessarily be the same as the enrichment of the bulk planet.   The interior density distribution of our own giant planets (where we have stronger constraints on mass, radius, gravity harmonics and rotation rate) remains poorly understood, and water in particular is locked away by deep condensation clouds on all four giants \citep[e.g.,][]{94fegley}.  As the principle carrier of oxygen, the third most abundant element in our solar system, water is crucial in the core accretion theory for the trapping and delivery of other volatile compounds to the forming planets, either as amorphous ices or as water-ice cages known as clathrate-hydrates \citep{05gautier}, and our inability to measure Jupiter's deep O/H ratio prevents us from establishing whether Jupiter has a solar-like C/O ratio of 0.54 \citep[e.g.,][]{04lodders}.  Thus the interpretation of planetary formation scenarios in our own solar system is non-unique, but the elemental and isotopic enrichments measured by the Galileo probe in Jupiter's $p<20$ bar region \citep{98niemann, 03atreya} suggest that any model assuming solar composition will be a poor choice \citep[e.g., see the review of Jupiter's origins by][]{04lunine}.  A planet would have a similar composition to its parent star only if the gravitational instability model \citep{97boss} were the dominant method of planetary formation in the system.  Core accretion theory predicts atmospheres enriched in heavier materials due to (i) erosion of the original core; (ii) gases accreted as H$_2$ and He were captured; and (iii) accretion of rocky and icy planetesimals during the later stages of evolution.

Recent suggestions of carbon-rich planetary compositions by \citet{12madhu} and \citet{13moses} raise interesting questions about compositional differences between O-rich accretion discs and C-rich ones.   O-rich silicates would be available to form protoplanetary cores in the former, to be displaced by carbides (e.g., SiC) and pure carbon minerals like diamond or graphite in the latter case \citep{10bond}.  Alternatively, carbon-rich planets would need to form in regions where carbonaceous materials dominated over water ice \citep{13moses}, such as `dehumidified' regions inward of the cold-trapping of water ice by the snow line \citep{88stevenson} or in locations where refractory organics or `tar' can be produced locally \citep{04lodders}.  In summary, the availability and composition of planetary building blocks depends on a broad variety of factors that are poorly understood even within our own solar system, but provide the source material for the subsequent chemistry shaping the atmospheres of the EGPs.  As a consequence, classification systems for EGPs should rely on direct compositional observables (e.g., elemental abundances and isotopic ratios), rather than model-dependant extrapolations to the epoch of planetary formation.

\subsection{Thermochemistry and Equilibrium}

To first order, the distribution of species in a planetary atmosphere can be understood in terms of thermochemical equilibrium, which is expected to work within the interior and deep troposphere.  The thermal structure determines the stability fields for different species, and as a planet ages and cools its atmosphere can pass through a number of transitions where a particular species acquires dominance.  Water is always expected to be one of the most abundant molecules, being stable under a wide variety of conditions \citep{99burrows}.  Higher temperatures break relatively weak bonds (such as C-H) in favour of molecules with stronger bonds \citep[such as C-O or C-N,][]{13moses}.  Focussing on the species with the highest cosmic abundances (H, He, C, O and N) that produce the most significant signatures in EGP spectra, high-temperature EGPs will favour CO and N$_2$ as the principle reservoirs of C and N, whereas cold EGPs will be in the CH$_4$ and NH$_3$ stability fields.  CO is favoured for temperatures exceeding around 1300 K (see Fig. \ref{exocloud}), and the CO abundance will increase exponentially at the expense of methane as we move further into the CO-stability field,  although the transition also depends on pressure \citep[e.g.,][]{03sudarsky}.  Similar stability arguments for sulphur, phosphorus, alkalis, halides and metal species should also be included to form an atmosphere in thermochemical equilibrium \citep[e.g.,][]{94fegley, 06visscher}.  

The precise details of a thermochemical equilibrium model are sensitive to the assumed thermal structure (pressure and temperature) and the bulk planetary composition.  In particular, the C/O ratio has a major effect on the stability fields for CO, methane and water \citep{02lodders}.  In oxygen-rich atmospheres, water is always abundant and CO is expected to be a major carbon-carrying species at high temperatures.  Carbon-rich atmospheres will have water as the dominant oxygen species at low temperatures, but CO will dominate at high temperatures (with some CH$_4$, hydrocarbons and nitriles) as water declines in importance \citep{13moses}.   These compositional differences could be observationally-constrained, prompting \citet{12madhu} to propose the bulk planetary C/O ratio as a second dimension, in addition to the stellar irradiance, in their 2-D categorisation scheme for EGPs.  CO is always a major constituent of hot EGPs, but water and CO$_2$ are more common for C/O$<1$, whereas CH$_4$, HCN and C$_2$H$_2$ are more common for C/O$>1$ \citep{12madhu, 13moses}.  This scheme has the benefit of encompassing the one-dimensional `oxygen-rich' case of \citet{08fortney}, who suggested the presence or absence of TiO and VO as strong UV/visible absorbers to explain the observations of stratospheric inversions on hot EGPs.  With limited oxygen, \citet{12madhu} suggests that TiO and VO cannot form and generate stratospheric inversions on carbon-rich EGPs.  However, this classification scheme is currently hampered by the extreme challenge of constraining the C/O ratio in any of the planetary atmospheres studied to date (see Section \ref{spectra}), and the need to explain the depletion of oxygen (the third most abundant element) from these planets.  

Equilibrium provides an excellent starting point for models, and is intricately tied to the condensation chemistry discussed in the next section, but in all but the most highly-irradiated EGPs it is unlikely to be a valid assumption for the atmospheric composition \citep{11moses}.  Condensation, vertical transport and mixing, photochemistry and UV destruction of molecular bonds at millibar-to-microbar pressures all cause the atmosphere to deviate substantially from equilibrium expectations \citep{10line, 11moses}.  Nevertheless, future observations of an ensemble of Jupiters will search for the balance of the principle species (H$_2$O, CH$_4$/CO, NH$_3$/N$_2$) and their products (oxides, hydrocarbons, nitriles) to understand how thermochemical equilibrium conditions vary from planet to planet.

\subsection{Condensation Chemistry}

The condensate cloud sequence predicted by equilibrium condensation theory, combined with the sedimentation of condensates to form layers \citep[`rainout', e.g.,][]{69lewis}, has proven successful in explaining the broad trends observed in MLTY brown dwarfs as a function of their effective temperatures \citep{99kirkpatrick, 05kirkpatrick, 11cushing}.  Thus condensation chemistry, whereby volatiles are released into the gas phase rather than being locked up as condensates beyond a certain temperature, is a natural parameter for use in EGP characterisation \citep{03sudarsky, 08fortney}, and a large body of literature has been devoted to this subject \citep[see the recent review by][]{13marley}.  However, the condensation sequence is a sensitive function of the bulk composition, temperature profile and strength of vertical mixing; the resulting cloud compositions can be mixed and extremely complex (including the effects of photochemically-produced hazes); and this sort of condensation chemistry is a poor approximation to our own solar system giant planets.  Equilibrium condensation predicts that the uppermost clouds of Jupiter and Saturn should be NH$_3$ ice, although this has never been determined spectroscopically \citep[aside from a few small, localised regions,][]{02baines},; the condensate clouds do not form in the locations predicted by condensation theory; and the bulk of Saturn's hazy atmosphere is due to photochemical species, rather than condensation clouds \citep{09west}.

\begin{figure*}[tbp]
\centering
\includegraphics[width=13cm]{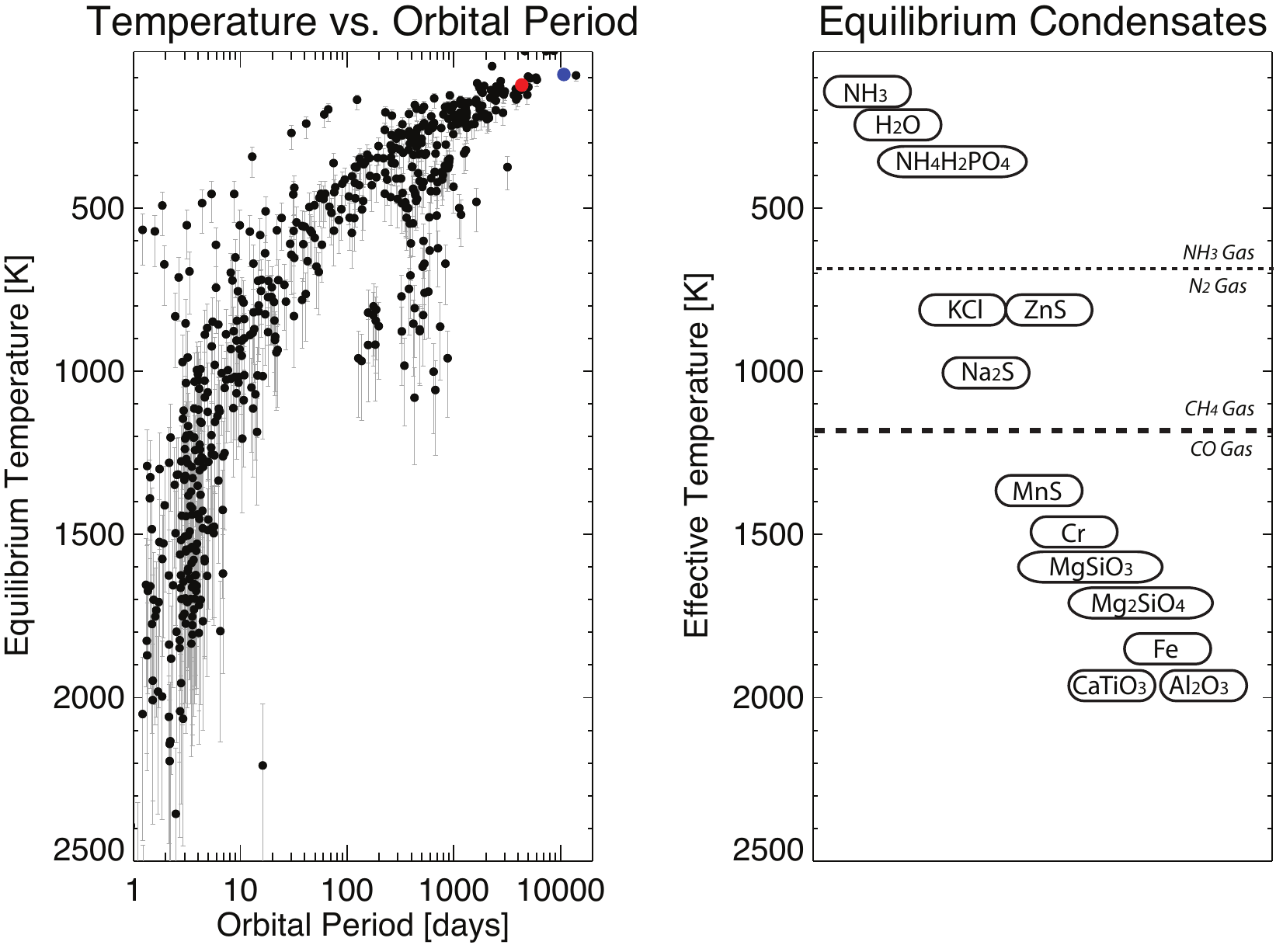}
\caption{The equilibrium temperatures of EGPs discovered to date (all data from the catalogue hosted at exoplanet.eu) compared to the expectations of thermochemical equilibrium cloud condensate schemes.  The left panel shows the relation between $T{eq}$ and orbital period, with the faster orbits providing a higher chance of characterisation via transit spectroscopy.  The right hand panel indicates the condensation effective temperatures (at 1 bar atmospheric pressure) for a range of potential condensates \citep{03sudarsky, 10lodders, 12morley}, assuming a solar-composition gas.  Dashed horizontal lines show the approximate stability regions for CO/CH$_4$ and N$_2$/NH$_3$ gas \citep{13moses}.  The coldest and most challenging Jupiters to observe as transiting planets are in the top right of panel (a), Jupiter and Saturn are included as the red and blue points, respectively.}
\label{exocloud}
\end{figure*}

The condensate sequence described by \citet{96fegley, 02lodders, 10lodders} and others provides a useful backdrop for the interpretation of brown dwarf, solar system and EGP spectroscopy, and one can imagine the full system of clouds being formed by condensation and buried deeper and deeper below the photosphere as the atmospheric $T(p)$ cools \citep{10burrows}, providing sinks for an increasing proportion of a planet's volatile species until only those with the lowest condensation temperatures remain in gaseous form to contribute to the emergent flux \citep[e.g., CH$_4$ only condenses in the frigid atmospheres of Uranus and Neptune,][]{04sanchez_clouds}.  The condensation of a cloud deck removes volatiles to alter the chemical equilibrium above the cloud \citep{96fegley} and modifies the reflective and absorptive properties of a planet's photosphere \citep[e.g.,][]{01ackerman}.  The condensation sequence is broadly described below \citep[following][and others]{03sudarsky}, and is indicated in Fig. \ref{exocloud}.

\begin{itemize}
\item \textbf{Hot Metallic Jupiters:} At the highest effective temperatures  ($T_{eff}\approx2500$ K, not to be confused with the equilibrium temperature, $T_{eq}$, which is the effective temperature in the absence of an intrinsic luminosity) all major rock-forming elements (Al, Ca, Ti, Mg, Si, Fe and Ni) are available in the gas phase, potentially leading to the formation of metal oxides (TiO and VO) which could generate stratospheric inversions \citep{03hubeny, 08fortney}.  These planets have been referred to as the `Pegasides' or `Roasters.'  Water, CO, Na and K are expected to dominate the spectrum \citep{10burrows}. The calcium, aluminium, titanium and vanadium gases are removed (so no longer contribute to any strong UV/visible absorption) as the first clouds begin to condense near $T_{eff}\approx1800-2200$ K, such as the refractory ceramics like corundum (Al$_2$O$_3$), calcium-aluminates like hibonite (CaAl$_{12}$O$_{19}$) or calcium-titanates like perovskite (CaTiO$_3$) \citep{10lodders}.   These hot cloud-free EGPs were the pM-class planets in the scheme of \citet{08fortney}.    

\item \textbf{Iron and Silicate Cloud Jupiters:}  As the atmosphere cools and the ceramics move down below the photosphere, pure iron begins to condense over the $T_{eff}\approx1500-2300$ K range, and as a cosmically-abundant species is likely to form thick cloud decks, removing metal hydrides (e.g., gaseous FeH condenses to Fe-metal, and CrH condenses to Cr$_2$O$_3$) from the gas phase by the time the planet has cooled to around $T_{eff}\approx1200-1500$ K \citep{03sudarsky}.  This homogeneous condensation and settling of a pure iron cloud, rather than allowing iron to consume the H$_2$S gas to form solid troilite (FeS), is the suggested reason for the presence of H$_2$S in the troposphere of Jupiter and Saturn \citep{96fegley, 89briggs}.   In a similar temperature range, the next clouds to form are the silicates in the $T_{eff}\approx1600-2000$ K range, removing magnesium- (e.g., Mg, MgOH, MgH) and silicon species (e.g., Si, SiO, SiS and silane, SiH$_4$) from the gas phase into the condensed phase \citep{94fegley}.    As the rock-forming silicates are more abundant than other metals, these form more substantial clouds of silicates such as forsterite (Mg$_2$SiO$_4$) and enstatite (MgSiO$_3$), two examples of the olivine and pyroxene sequences of silicates, although the precise combination of Mg, Si and oxygen is likely to be rather complex.  Forsterite will condense at the highest temperatures, so will be the likely consumer of most of the magnesium \citep{03sudarsky}.  Enstatite condenses at the lowest temperature for the silicates, so by the time this forms, all the major rock forming elements have been removed from the gas phase.  Note that this silicate condensation is an important oxygen sink, removing up to 20\% from the gas phase \citep{10lodders}.  The absence of gaseous SiH$_4$ on our giant planets could be evidence for trapping in a silicate cloud deck at great depth \citep{94fegley}.  These silicate cloud Jupiters were the hottest (class V) discussed by \citet{03sudarsky} and the cloudy-case (pL) discussed by \citet{08fortney}.  

\item \textbf{Sulphide Cloud Jupiters:}  The pressure-broadened wings of the gaseous sodium and potassium resonance lines dominate the visible spectra of some brown dwarfs and EGPs down to $T_{eff}\approx700-1100$ K \citep{03sudarsky,10lodders}, until they too are locked away by reactions to form Na$_2$S, NaCl, KCl and other oxides, hydroxides and hydrides, before being buried in condensed phases as sulphide, choride or halide clouds.   Because of their relatively low cosmic abundances, such clouds are likely to be thin and cirrus-like \citep{03sudarsky}, compared to the extensive silicate and iron clouds.  \citet{12morley} found that the emergence of a Na$_2$S cloud was the likely explanation for the enhanced cloudiness of cooler T dwarfs at around 600 K, rather than ZnS or MnS clouds which are limited by the low manganese and zinc abundances in a solar composition atmosphere.  Other tenuous clouds may also be possible at these temperatures, such as chlorides of Cs and Rb \citep{99lodders} or NH$_4$H$_2$PO$_4$, one of the condensates expected to remove PH$_3$ from the gas phase on Jupiter and Saturn \citep{94fegley}.

\item \textbf{Water Cloud Jupiters:}  As $T_{eff}$ drops still further, these clouds are buried beneath the photosphere until the point where the volatile species H$_2$O (250-350 K) and NH$_3$ (100-200 K) condense homogeneously to form the familiar cloud-decks of our cool Jupiter \citep[Class I and II in the scheme of][]{03sudarsky}.  With an exposed water cloud dominating the photosphere, a water-class jovian would have a bright reflective spectrum and H$_2$O would be removed as the dominant gas phase species in the spectrum.  CH$_4$ and NH$_3$ will have grown in importance as the temperature moved into their thermochemical stability realms at the expense of CO and N$_2$.    It is worth reiterating the fact that extensive NH$_3$ and H$_2$O cloud decks are yet to be verified in our own solar system, save for a few regions of strong convective activity, and that the measured altitudes of the main cloud decks do not agree with the predictions of equilibrium cloud condensation theory \citep[e.g.,][]{04west, 09west}.  
\end{itemize}

Although this sequence is certainly attractive, and is successful when applied broadly to brown dwarfs \citep[such as the waning importance of metallic oxides and the changing cloudiness at the MLT transitions,][]{12burgasser}, the details are the subject of considerable debate.  A planet will likely transition between these phases as they age and cool, our own Jupiter probably starting out in the $T_{eff}\approx600-1000$ K range before the alkalis and water were buried in the deeper troposphere \citep{10burrows}.  There are different modelling approaches for the formation of clouds at particular altitudes, and different treatments of the balance between vertical uplift, sedimentation and transport to supply cloud material and homogeneous/heterogeneous nucleation, evaporation and coagulation of cloud materials once they are in place  \citep[e.g.,][]{01ackerman, 01allard, 08helling}.  These differing assumptions can lead to substantial variations in the grain sizes, densities and altitudes between the models.  Some clouds require a chain of chemical reactions to form, rather than simple condensation (e.g., NH$_4$SH on Jupiter and Saturn).  Homogeneous nucleation requires high supersaturations not typically observed in planetary atmospheres, whereas heterogeneous nucleation requires activated cloud condensation nuclei to be preexisting before condensation occurs - see \citet{13marley} for a review of the complexities of cloud formation.  

More significant drawbacks are the assumptions about the internal structure (a cool interior may lock species away far below the photosphere), temperature profile (crossing the cloud-condensation curves multiple times can provide multiple cold traps for different species) and bulk composition (often assumed to be solar, sometimes with enhanced metallicity), all of which would confuse the sequence of condensate clouds outlined above.  Based on our experience in the solar system, we would expect a veritable zoo of species (both condensible volatiles and photochemically-produced hazes) to mix together to form clouds at altitudes vastly different to the expectations of equilibrium.  As pointed out by \citet{13marley}, clouds are likely to be the `limiting factor' in our ability to understand EGP spectra, but they must play a crucial role in any EGP categorisation scheme \citep{03sudarsky, 08fortney, 10lodders}.   Finally, the clouds and hazes often provide broadband contributions to emergent spectra without spectral features to permit an unambiguous detection.  However, future spectroscopic characterisation of an ensemble of Jupiters could show the temperatures at which different species are released from the condensate into the vapour phase (e.g., TiO, VO, CrH, FeH, SiH$_4$, SiS, Na, K, etc.), allowing indirect identifications of the cloud-forming species.

\subsection{Atmospheric Mixing and Redistribution}

The equilibrium expectations of thermochemistry and condensation chemistry could be radically altered if significant vertical and horizontal redistribution of material occurs, mixing condensates and gas phase species together.  Transport-induced quenching has been studied in one-dimensional models \citep{10line, 11moses, 13moses}, where a gas parcel rising to cooler altitudes transports species `frozen in' at abundance ratios pertinent to deeper, higher pressures (their quench points).  Above the quench point for a particular species, the transport and mixing timescale (parameterised by a vertical eddy mixing coefficient, $K_{zz}$) is faster than the chemical kinetic timescale for reactions depleting the gas, so the species is enhanced over equilibrium expectations \citep{84lewis}.  Phosphine, CO, germane and arsine are good examples of disequilibrium species in the tropospheres of Jupiter and Saturn which would not be observable without transport-induced quenching \citep[e.g.,][]{09fletcher_ph3, 11fletcher_vims}.  An overabundance of CO due to transport-induced quenching has also been established in T dwarfs \citep{07saumon}.  

For irradiated EGPs, \citet{11moses} demonstrated that quenching was important for the abundances of methane, ammonia and HCN, which are disequilibrium species when temperatures are within the CO and N$_2$ stability fields, although at the highest temperatures the chemical-kinetic reactions are typically fast enough to return the composition to equilibrium.  CO, water and N$_2$ also quench, but this would only be important for cooler jovians within the CH$_4$ and NH$_3$ stability fields \citep{11moses}.   The predicted abundances of these disequilibrium species depends on the equilibrium composition at the quench point, and the altitude of the quench point is sensitive to (i) the modellers' chosen value of $K_{zz}$ (e.g., a deeper quench point would cause larger quenched abundances for methane and ammonia); and (ii) uncertainties in the kinetic reaction rates (see the paper by J. Moses \textit{et al.} in this volume).  Furthermore, the transition between reaction-timescale-dominated to transport-dominated may be rather gradual, complicating precise predictions of species abundances.  Alternatively, abrupt quenching can result if thermal gradients are large as reaction rates are often exponential functions of temperature.  So in order to use any molecular composition as a dimension of a EGP classification scheme, one must understand the potential sources and sinks at the level of interest, requiring good estimates of the strength of vertical mixing ($K_{zz}$) from dynamical models.

\citet{10showman} provide an excellent introduction to the broad field of EGP dynamics, and the hierarchy of models required to determine the fundamental processes governing atmospheric mixing.  The dynamic state of a planetary atmosphere is a sensitive function of the rotation rate, which can be used to categorise planets in different circulation regimes \citep{11read}.  However, as the rotation rates of EGPs will remain uncertain for all but tidally-locked planets, and as they do not significantly affect the emergent spectrum, we will not consider them further.  However, the emergent spectrum can be influenced by the efficiency of redistribution of irradiation, which depends on the proximity of the convective zone to the photosphere (the region of radiative cooling to space).  On Jupiter and Saturn, the radiative-convective boundary is within the photosphere, so that the convective motions can uniformly redistribute heat around the planet \citep{90ingersoll} to ensure that these planets emit near-isotropically in the thermal infrared.  But if the irradiance cannot penetrate the convective zone (which moves to deeper, high pressures as a function of increasing irradiance), then heat is no longer efficiently redistributed, leading to strong thermal contrasts as a function of longitude on tidally-locked EGPs \citep{07knutson}, possibly perturbed by zonal winds in the radiative regions.  The same is true for composition if chemical lifetimes are shorter than transport timescales, producing compositional contrasts from dayside to nightside rather than homogenising the composition with longitude via transport-induced quenching.  Such contrasts complicate the combined analysis of transmission spectra (sensitive to the planetary limb/terminator) and emission spectra (sensitive to the dayside) of tidally-locked EGPs.   The most highly-irradiated EGPs are expected to have the least efficient convective redistribution of energy.

The atmospheric opacity due to gases and clouds determines the depth of penetration of insolation, and in turn determines the efficiency of energy and material redistribution.  Combining radiative transfer, chemistry and circulation models could provide a better handle on how global three-dimensional processes affect disc-averaged spectra, but it is clear that the efficient transport-induced quenching and horizontal mixing will negatively impact attempts to create a compositional classification scheme for EGPs.

\subsection{Photochemical Disequilibrium}

The final processes considered here are the photolytic reactions excited by absorption of UV photons from the parent star, dissociating molecules to form new disequilibrium products and driving the composition away from equilibrium \citep{04liang, 10line, 11moses, 13moses}.  The rate of photolysis depends on: (i) the availability of species released from the condensed phase and provided by transport-induced quenching; (ii) the shielding properties of high-altitude aerosols and hazes, which can modify photolysis rates; and (iii) the stellar flux in key photolytic bands, which in turn depends on the stellar type (e.g., different stellar spectra from M to F in Fig. \ref{Teqm} could excite different photolytic pathways).  In their survey of highly-irradiated Jupiters, \citet{11moses, 13moses} find that methane and ammonia (enhanced in the photosphere by transport-induced quenching) are depleted by photolysis in favour of atomic species (especially hydrogen), unsaturated hydrocarbons (particularly C$_2$H$_2$), nitriles (HCN) and some radicals.  The opacity of these species, particularly HCN and C$_2$H$_2$, could have a significant impact on EGP spectra, and yet they are rarely incorporated into spectral models \citep{12madhu}.  If photochemistry dominates in the stratosphere, as it does on the giant planets of our own solar system \citep[e.g.,][]{05moses_jup}, then we may even see the formation of C$_3$H$_x$ compounds at the highest altitudes (potential haze and soot precursors), and the formation of benzene and complex nitriles on the nightside \citep[they never exceed parts-per-billion levels on the illuminated dayside,][]{11moses,13moses}.  Indeed, cooler EGPs would have a greater likelihood of forming these complex polyaromatic hydrocarbons and nitriles than the hottest EGPs.  In addition, ion chemistry could potentially lead to the formation of complex photochemical haze precursors, as it does on Titan.  CO, H$_2$O, N$_2$ and CO$_2$ are relatively unaffected by this disequilibrium chemistry due to either their strong bonds or efficient recycling \citep{11moses}.  At higher temperatures, particularly when all species are released into the gas phase, complex hydrocarbons and nitriles will have difficulties surviving, the high temperatures favouring a return to thermochemical equilibrium \citep{11moses}.

The picture emerging is one of a transition from photochemically-rich photospheres at low temperatures, where disequilibrium chemistry can dominate over the equilibrium conditions, to photochemically-poor photospheres at the highest temperatures favouring kinetic equilibrium.  But the photochemical models are extremely sensitive to the thermal structure; shielding from atmospheric aerosols and strength of vertical mixing; and are lacking some species \citep[e.g., sulphur and phosphorus,][]{09zahnle,06visscher} which may play important roles.  For example, the photolytic products hydrazine (N$_2$H$_4$) and diphosphene (P$_2$H$_4$) are expected to be the principle contributors to the hazes in the upper tropospheres of Jupiter and Saturn, although these have never been spectrally identified by remote sensing.  Furthermore the disequilibrium models have proven incapable of reproducing the high CO$_2$ abundances inferred on several EGPs \citep{13moses, 12line, 12lee, 09madhu}, either due to missing species in the spectral retrievals or missing photochemical pathways in the model.  

\citet{10knutson} propose an EGP categorisation scheme in terms of the level of chromospheric activity in the parent star, suggesting that photolytic processes break down the UV/optical absorber responsible for the stratospheric inversions in the scheme of \citet{08fortney}.  The more active the host star, the less likely a planet is to form a stratospheric inversion, irrespective of the atmospheric temperature.  This requires more detailed investigation, as extreme UV flux, ion chemistry and charged particle bombardment may instead generate more photochemical `smog' in the upper atmosphere, which could in turn be a localised source of heating to form stratospheric inversions \citep{11moses}.  Nevertheless, the apparent correspondence between the stellar activity and the planetary spectrum may have intriguing consequences for EGP categorisation schemes.

\section{Spectral Retrieval and Interpretation Challenges}
\label{spectra}

The previous section reviewed the processes governing the chemical composition of an EGP's observable photosphere - the bulk composition and metallicity imprinted at the time of planet formation; thermochemical equilibrium; the sequence of condensate cloud formation; atmospheric mixing and transport-induced quenching; and photolytic formation and destruction.  Each process is intricately linked in a myriad complex ways (Fig. \ref{cartoon}), but at least permits us to understand the potential range of atmospheric compositions soon to be measured in the ensemble of EGPs.  Multi-dimensional classification schemes have been proposed relying on the strength of stellar irradiation (parameterised as $T_{eq}$); the condensate sequence by analogy to the MLTY brown dwarf sequence \citep{02lodders, 03sudarsky, 08fortney}; the atmospheric chemistry and balance between species such as carbon and oxygen \citep{12madhu, 13moses}; and the effects of extreme stellar activity on upper atmospheric composition \citep{10knutson}.  Of all these, only $T_{eq}$ is model-independent, which is a poor proxy for the true atmospheric temperatures.  All of these processes will be active in EGP atmospheres, such that Jupiter-class objects exist as a continuum of different `types.'   Indeed, planets will typically move along this continuum as they age and cool, different condensates are buried beneath the photosphere, and new gaseous species come to dominate the chemistry and haze formation.  This shifting between classifications could also be caused by strong day/night contrasts, or eccentric orbits \citep{08fortney}.

Compositional classifications are more complex than the spectral classifications employed for brown dwarfs \citep[i.e., identifying the $T_{eff}$ when a molecular band becomes visible in a spectrum,][]{05kirkpatrick}, but in the presence of strong irradiation they are arguably more informative.  Although formational, dynamical and chemical models have a strong sensitivity to the initial assumptions, the key parameter plaguing EGP characterisation is the absence of reliable observational constraints for an ensemble of Jupiter-class objects, as outlined below.  

\subsection{Degeneracies in Spectral Retrieval}
Modelling of EGP transit and eclipse spectroscopy has fallen into three broad categories - forward modelling millions of spectra based on a variety of assumptions about the global temperature structure, clouds and composition and then finding the subsection of phase space which best fits the data \citep[e.g.,][]{05seager, 06burrows, 07burrows, 06fortney, 08fortney, 09madhu}; a Markov Chain Monte Carlo (MCMC) exploration of phase space, accepting the random-walk steps that improve the fit to the data \citep{11madhu,12benneke}; or employing an iterative retrieval scheme to find the family of statistically-plausible solutions consistent with the available data \citep[e.g., optimal estimation, ][]{00rodgers,08irwin,12lee,12line,13barstow,13line}.  In the latter cases, the intention is to remove bias to any particular assumptions (C/O ratio, solar composition, equilibrium versus disequilibrium) and investigate the possibilities supported by the measurements in as large a parameter space as possible.  Interpretation of the resulting temperatures, hazes and composition via models can then be used to narrow down the parameter space, but different authors disagree on the details of the application of different retrieval architectures (optimal estimation, constrained linear inversion, MCMC techniques, nested sampling algorithms, etc.) and cloud models \citep[e.g., scattering versus non-scattering aerosols,][]{12dekok}.

In the three cases described above (forward modelling, MCMC and optimal estimation), the exploration of parameter space is likely to be incomplete, as it relies on a complete knowledge of the prior to allow a model to `roam' over all possible combinations of parameters.  At best, spectral retrieval techniques \textit{bracket reality} from within a subsection of parameter space.  The range of parameter space to be explored is continuously evolving in tandem with data improvements (e.g., the addition of new techniques for parameterising aerosol and cloud influences on photospheric spectra; inclusion of updated spectroscopic line parameters, etc.).  The key benefit of a spectral retrieval is that it aims to be data driven, rather than model driven, to identify the potentially-vast region of parameter space consistent with the data.  In some cases, it shows that even high-quality data can reveal very little about the atmosphere in question, and highlights the importance of \textit{broad spectral coverage} to constrain the shape of the spectrum from the ultraviolet to the infrared.  In the solar system, spectral retrieval is a useful technique for rapid determinations of atmospheric properties from huge numbers of spatially-resolved spectra.  Although this is not currently the case for exoplanets, the number of transit spectra (and hopefully directly-imaged spectra) available is set to increase dramatically in the coming decades due to planned ground-based and space-based observing programs, requiring a computationally-efficient technique for analysis and cross-comparison of data from a large number of different planets.

Spectral retrievals are poorly constrained if the number of atmospheric parameters exceeds the available measurements, which is the case for the majority of EGP spectra available today.  With only a handful of broad filter-averaged measurements, a wide range of plausible atmospheres are permissible, often so broad that we cannot hope to classify the atmosphere with any certainty (for example, the prevalence of stratospheric inversions on EGPs is still in doubt).  Simplifying assumptions, such as an aerosol-free atmosphere, are necessary but potentially misleading \citep[e.g.,][]{12pont,12lee}, and there remains significant degeneracy between temperatures, aerosols and gaseous composition to plague the study of these planets.  The spectral models used by \citet{12madhu} in their C/O classification scheme are aerosol-free, for example.  All this implies that retrieved parameters carry large, often insurmountable, uncertainties which restrict their usefulness.

Clouds and hazes present a particular problem that remains unresolved in our own solar system - the spectral signatures of silicate clouds, iron clouds, sulphide/chloride clouds, and water/ammonia ice clouds are typically broad and flat, responsible for shaping the continuum from the UV to the infrared and reducing peak-to-trough contrasts in molecular bands \citep[e.g.,][]{03sudarsky}.  Non-spherical aggregates, wide ranges of particle size distributions and mixtures of condensate and photochemical particulates all serve to mask spectral signatures, such that NH$_3$ ice on Jupiter has only been identified in regions of strong convective updrafts \citep{02baines, 10sromovsky}.  Deducing atmospheric properties from narrow spectral ranges (e.g., from a single instrument) would be misleading; only by combining multiple data points to form a broad-band low resolution spectrum can we hope to understand these aerosols.  Alternatively, for cooler EGPs we could exploit both the reflection and emission components from the cloud decks to deduce their properties (light reflected from clouds and hazes dominates the spectrum below 4-5 $\mu$m on Jupiter and Saturn).

\subsection{Challenging Data}
The capabilities of EGP remote sensing have improved tremendously in recent years, but in most datasets the instrumental systematics dwarf the expected transit signal, requiring sophisticated (and sometimes controversial) decorrelation techniques to extract the measurements \citep[e.g.,][]{11gibson,12waldmann}.  Furthermore, to create broad-band spectra from the UV to the infrared we must stitch together non-simultaneous measurements, sometimes only marginally compatible with one another \citep[e.g.,][]{12lee,12pont} due either to global variations on the planet itself (unlikely given the experience of jovians in our own solar system); peculiarities of the instrument used for the observations; or due to the variability of spots on the parent star masquerading as different transit depths.  One promising ground-based technique to break the degeneracies between temperature and composition is to use high spectral resolutions to unambiguously detect molecular bands and the Doppler shift of lines with orbital phase to separate the planet signal from the stellar and terrestrial background \citep[e.g.,][]{10snellen, 12brogi, 13dekok}.  Space-based spectroscopy in the coming decades from observatories dedicated to transit spectroscopy (i.e., improving on the work of Hubble and Spitzer); coupled with spectroscopy of directly imaged planets \citep[e.g., spectroscopy of HR8799b,][]{13lee}, should help to rapidly expand the ensemble of EGPs available for the testing of classification systems.  

A spectral retrieval model will only be as good as the line database used, and the high temperatures (up to 2500 K) for highly irradiated EGPs is a significant challenge for experimental work.  Some success has been provided by theoretical calculations of molecular spectra \citep[water, for example, is now included in the HITEMP spectral database,][]{10rothman,06barber}, but the opacities due to methane, HCN, C$_2$H$_2$, TiO and VO (among others), and information on the pressure-broadening of Na and K lines \citep{00burrows} remain inadequate to disentangle their competing effects in EGP spectra.  In the absence of high-resolution EGP spectra, the identification of unique bands for these species is challenging, leading to their omission from some spectral models.  Indeed, the uncertainties on transit spectra are so large that the inclusion of additional atmospheric species is sometimes not warranted to fit the data within the error \citep{12lee}.  In other words, the current data allows EGP composition to range over many orders of magnitude and still fit the measurements.  Only as new high-temperature datasets become available, and transit measurements improve, will we be able to make progress in classifying the compositions of these EGPs.

\section{Conclusion:  An Ensemble of Jupiters}

This review has demonstrated that the Jupiter-class of EGP represents a broad continuum of planetary types, with bulk composition, equilibrium and disequilibrium chemistry, and the complex sequence of condensates all competing to shape the emergent spectra.  Classification schemes involving stellar irradiance and condensate formation, bulk composition and chemistry, and the influence of strong stellar activity have all been proposed \citep{03sudarsky, 08fortney, 10knutson, 12madhu, 13moses} and capture the essence of the EGP classification problem.  All these schemes are model-dependent, and they are all hampered by the lack of observational constraints, particularly for the cooler jovians bridging the gap between our Solar System and the two `Rosetta Stones', HD 189733b ($T_{eq}=1200_{-105}^{+230}$ K, a hazy Jupiter lacking a thermal inversion, orbiting a cool chromospherically-active K star) and HD 209458b ($T_{eq}=1440_{-125}^{+270}$, a cloud-free, hot metallic Jupiter with a thermal inversion orbiting a G star).  

Classification requires a robust ensemble of atmospheric compositional types, under varying irradiation conditions (i.e., different stellar types, power and orbital radii).  It also requires spectroscopy that is both accurate and sufficiently broadband to (i) determine the continuum formed by atmospheric temperatures and hazes and (ii) unambiguously detect the presence of molecular species.  Space-borne spectroscopy from the UV to the infrared, potentially from JWST or ECHO \citep{12tinetti}, could begin to validate some of the early findings on irradiated EGPs and extend coverage to planets with longer orbital periods and smaller stellar influences (see Fig. \ref{exocloud} for the range of $T_{eq}$ within reach of such transit studies).  Based on the current statistics of targets suitable for EChO, up to 50\% of the mission time could be devoted to the Jupiter class \citep{12tinetti}.  Hot metallic Jupiters and silicate cloud Jupiters will be well-sampled across all stellar types (the \textit{hot} sample, $T_{eq}>1800$ K), but sulphide cloud Jupiters, cloud free Jupiters with strong alkali lines ($700<T_{eq}<1800$, the \textit{temperate} sample) and water-cloud jovians (the \textit{cool} sample) could also be targeted.  Cooler Jupiters must be observed around cooler M and K stars, so that their orbits are still sufficiently short to permit the observation of multiple transits (Fig. \ref{exocloud}).  EGP categorisation schemes will allow us to make optimal selections of targets for such a mission, and only by assembling a reference collection of well-characterised EGP photospheres for a range of different stellar flux conditions can we begin to test the predictions of the EGP schemes reviewed here.  We may find that certain planets could be considered archetypes of their classes, in the same way as Jupiter is seen for our solar system giant planets.  Ultimately it seems certain that understanding this `continuum of jovians' will lead us to view our own cold giant planets in an entirely new way.

\section*{Acknowledgements}
Fletcher is supported as a Royal Society Research Fellow at the University of Oxford.  The UK authors acknowledge the support of the Science and Technology Facility Council.  Barstow is supported by the John Fell Oxford University Press (OUP) Research Fund. The authors thank J. Moses, N. Madhusudhan, M. Line, N. Gibson and F. Pont for enlightening discussions.
%\appendix{}   

%\section*{Tables}

%   \begin{table}
%   \caption{<title>}
%   \longcaption{<long description>}
%   \begin{tabular}{<column alignments}
%   \hline
%   <headings>\\
%   \hline
%   <body of table>
%   \hline
%   \end{tabular}
%   \end{table}

%\begin{thebibliography}
%\end{thebibliography}
%\clearpage

\bibliographystyle{harvard}
\bibliography{../../references_master}

\label{lastpage}
\end{document}